\documentclass[aps,pra,twocolumn,superscriptaddress]{revtex4-1}
\usepackage{graphicx}
\usepackage{amsmath}
\usepackage{nccmath}
\usepackage{amsfonts}
\usepackage{xcolor,colortbl} 
\usepackage{pifont}
\newcommand{\cmark}{\ding{51}}%
\newcommand{\xmark}{\ding{55}}%
\definecolor{Gray}{gray}{0.85}
\definecolor{LightCyan}{rgb}{0.88,1,1}
\definecolor{camel}{HTML}{FFCC99}
\newcolumntype{a}{>{\columncolor{Gray}}c}
\newcolumntype{d}{>{\columncolor{white}}c} 
\begin{document}
\title
{Scattering of Spin-$\frac{1}{2}$ Particles from a $\cal PT$-symmetric Complex Potential}
\author{Ege \"Ozg\"un} 
\email{ozgune@bilkent.edu.tr}
\affiliation{NANOTAM-Nanotechnology Research Center, Bilkent University, 06800 Ankara, Turkey}
\author{T. Hakio\u{g}lu} 
\affiliation{Energy Institute and Department of Physics,\.{I}stanbul Technical University,34469 \.{I}stanbul, Turkey}
\affiliation{Department of Physics, Northeastern University, Boston, MA 02115 , USA}
\author{Ekmel Ozbay} 
\affiliation{NANOTAM-Nanotechnology Research Center, Bilkent University, 06800 Ankara, Turkey} 
\affiliation{Department of Physics, Department of Electrical and Electronics Engineering and UNAM-Institute of Materials Science and Nanotechnology, Bilkent University, 06800 Ankara, Turkey}

\begin{abstract}
In this letter, we study the scattering of spin-$\frac{1}{2}$ particles from a spin-independent parity time ($\cal PT$)-symmetric complex potential, and for the first time, theoretically demonstrate the coexistence of $\cal PT$-symmetric and $\cal PT$-broken phases for broadband energy spectra in this system. We also show the existence of anisotropic transmission resonances, accessible through the tuning of energy. Our results are promising for applications in spintronics, semiconductor-based devices, and a better understanding of the topological surface states.
\end{abstract}

\maketitle
\section{Introduction}
With their gamechanger paper published in 1998 \cite{qm1}, Bender and Boettcher questioned the condition of Hermiticity as a seventy-year-old conceptual foundation in quantum mechanics and proposed that \cite{qm1,qm2} it should be replaced by the $\cal PT$-symmetry. In a series of three manuscripts \cite{mzade1}, Mostafazadeh introduced the concept of pseudo-Hermiticity and showed that every Hamiltonian with real spectra is pseudo-Hermitian and that the $\cal PT$-symmetric Hamiltonians all belong to that class of pseudo-Hermitian Hamiltonians. Later, in the light of $\cal PT$-symmetric quantum mechanics, thanks to the analogy between Schr\"odinger's equation and the equation for the propagation of an electromagnetic wave under paraxial approximation, $\cal PT$-symmetry studies were ignited in the field of classical optics \cite{El-Ganainy, Musslimani, Makris1, Klaiman, Guo, Makris2, Ruter, Ramezani}. The studies on optical systems with equal loss/gain media that show $\cal PT$-symmetric nature, gave rise to the derivation of generalized unitarity relation \cite{Ge}, also coined as the pseudo-unitarity condition, which was studied later on within the context of quantum mechanical scattering \cite{Ahmed,mzade2}. Although there is an ocean of significant theoretical studies on $\cal PT$-symmetric structures, the experimental proposals and demonstrations are far more limited. Observation of \cal $\cal PT$-symmetry breaking in optical systems \cite{Guo,Ruter}, $\cal PT$-symmetry in optically induced atomic lattices \cite{Zhang}, in a single quantum system \cite{Wu}, with superconducting quantum circuits \cite{Tan}, experimental realization of Floquet $\cal PT$-symmetric systems \cite{Chitsazi}, demonstration of optical anti-$\cal PT$-symmetry in a warm atomic-vapour cell \cite{Peng}, applications of $\cal PT$-symmetry in optics including proposals and demonsrations of lasing \cite{Feng}, cloaking \cite{cloak1, cloak2} and uni-directional invisibility \cite{Lin}, and proposals for experimental realizations within the context of condensed matter physics \cite{Wunner} and atomic gases \cite{Hang} are among those. There are also reviews \cite{rev1,rev2,rev3} and books \cite{book1,book2} covering all of the details of the subject that are excluded within the scope of this letter.

\begin{figure} [t]
\includegraphics[scale=0.4]{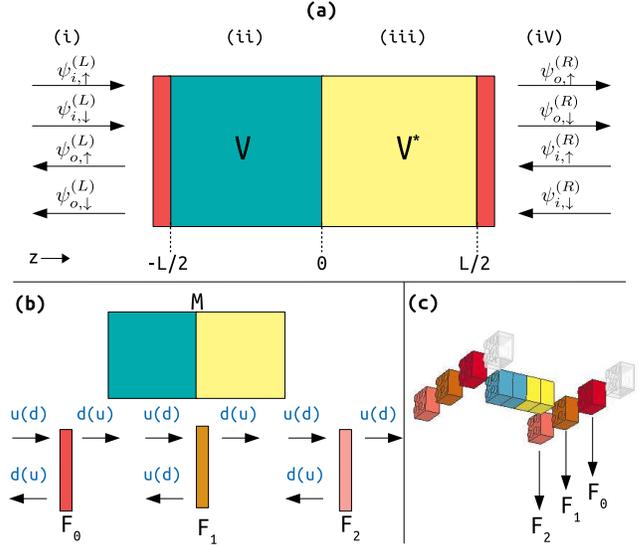}
\caption{{\bf (a)} Overview of the structure {\bf (b)} The abilities of symmetric ($F_0$) and asymmetric ($F_1$ and $F_2$) spin flippers (SF), (F=L/R for left/right). u(d) denotes up(down) spin states. {\bf (c)} The schematic of 16 possible configurations, where the transparent component means not using any SF.}
\label{structure}
\end{figure}
Recently, a photonic heterostructure with one-dimensional gain/loss bilayer and polarization converting components, which allowed $\cal PT$-symmetric and $\cal PT$-broken eigenvalues to coexist simultaneously for broadband wavelengths, was studied \cite{ozgun-andriy}. In this paper, through a simple yet intuitive problem of one-dimensional scattering of spin-$\frac{1}{2}$ particles from a spin-independent, complex, non-Hermitian $\cal PT$-symmetric potential, we study the quantum mechanical analog of that photonic problem and theoretically demonstrate the mixing of $\cal PT$-symmetric and $\cal PT$-broken eigenvalues of the scattering ($\mathbb{S}$) matrix. To the best of our knowledge this problem has also not been studied in the context of quantum device applications. The possibility of obtaining mixed phase is already giving rise to intriguing research in the field of photonics \cite{Kafesaki}, which makes our results even more important, since we are expecting the onset of similar research in the field of quantum mechanics.

\section{Proposed Structure and Theory}
\label{s2}
Our system, as given in Fig.~\ref{structure}, consists of a $\cal PT$-symmetric complex potential region (M), with $V=V_R+iV_I$, $[V_R, V_I \in \Re]$ satisfying the $\cal PT$-symmetry condition $V(z) = V^{\ast}(-z)$ and three different types of spin flippers (SF), that create a four-channel system -two input and two output channels for up (u) and down (d) spins that can be attached to both ends of M. Various proposals for such $\cal PT$-symmetric structures are given in the introduction part of this manuscript and is not our main focus. Incoming and outgoing wavefunctions of the spin-$\frac{1}{2}$ particles are shown in Fig.~\ref{structure}(a). The SFs flip the spin of the reflected and transmitted spin-$\frac{1}{2}$ particles in the following way as shown in Fig.~\ref{structure}(b): The symmetric SF $F_0$ flips the spin of both reflected and transmitted spin-$\frac{1}{2}$ particles from u(d) to d(u), whereas the asymmetric SF $F_1$ only flips the spin of the transmitted ones and $F_2$ does the opposite and only flips the spin of the reflected ones. The SFs are heterojunction interfaces based on materials with strong spin-dependent potentials operating on a certain incoming spin state $\vert S_i\rangle=a\vert\uparrow\rangle+b\vert\downarrow\rangle$. Their basic function is to manipulate the Bloch-sphere of the transmitted $\vert S_t\rangle$ and the reflected $\vert S_r\rangle$ states by external parameters like the electric field vector $\bf E$ and the controllable gates that can be tuned externally \cite{spin-prism1}. Moreover, it was theoretically demonstrated that it is possible to obtain strongly spin-flipping resonances, which are required for the SFs proposed in this letter, in quantum well structures by exploting the Rashba and Dresselhaus spin orbit couplings \cite{spin-prism2}. III-V group semiconductor heterostructure quantum wells with sizes ($10-70 \; nm$) proposed in Ref. \cite{spin-prism2} can be fabricated using conventional epitaxial techniques, such as Molecular Beam Epitaxy, Atomic Layer Deposition, and Metalorganic Chemical Vapour Deposition \cite{nanofab}.
Here, we assume that the SFs only manipulate the spin degree of freedom of the incoming spin-$\frac{1}{2}$ particles and not the overall reflection-transmission profile, which provides a simplification in calculations. We can safely do this, since our main results, that are mixing of $\cal PT$-symmetric and $\cal PT$-broken phases and the existence of anisotropic transmission resonances (ATR), would still be valid without the assumption.

We relabel the SFs $(F)$ as $R$ and $L$ for right and left placement, so that we have three different possibilities for right placement $R_0, R_1, R_2$, and three for left placement $L_0, L_1, L_2$. Also taking into account that we can choose to not use any of those for right and left and only stick with $M$ (illustrated with the transparent component), we have 16 different possibilities in total, summarized in the schematic given in Fig.~\ref{structure}(c). Let us first lay the physics of the problem before investigating these possible configurations in detail.

The most general solution for the Schr\"odinger's equation in regions with (ii,iii) and without (i,iv) the $\cal PT$-symmetric complex potential $V=V_R+iV_I$ in Fig.~\ref{structure} are given by: $\psi(z) = Ae^{ik_0z}+Be^{-ik_0z} $ for $V=0$ and $\psi(z) = Ce^{ik_1z}+De^{-ik_1z}$, for $V\neq0$, where $k_0=[2mE/\hbar^2]^{1/2}$, $k_1=[2m(E-V)/\hbar^2]^{1/2}$, in which $m$ is the mass of the spin-$\frac{1}{2}$ particle, $E$ denotes the energy and $A,B,C$ and $D$ are to be determined by boundary conditions. For the four different regions given in Fig.~\ref{structure} we can write the wavefunctions of the spin-$\frac{1}{2}$ particles and then use the appropriate boundary conditions to find reflection coefficients (from left and right) $r_L$, $r_R$ and the transmission coefficient $t$, which in return gives right reflectance $R_R= \vert r_R \vert^2$ left reflectance, $R_L = \vert r_L \vert^2$ and transmittance $T = \vert t \vert^2$, which obey the pseudo-unitarity or in other words the generalized unitarity relation \cite{Ge}:

\begin{figure} [t]
\includegraphics[scale=0.85]{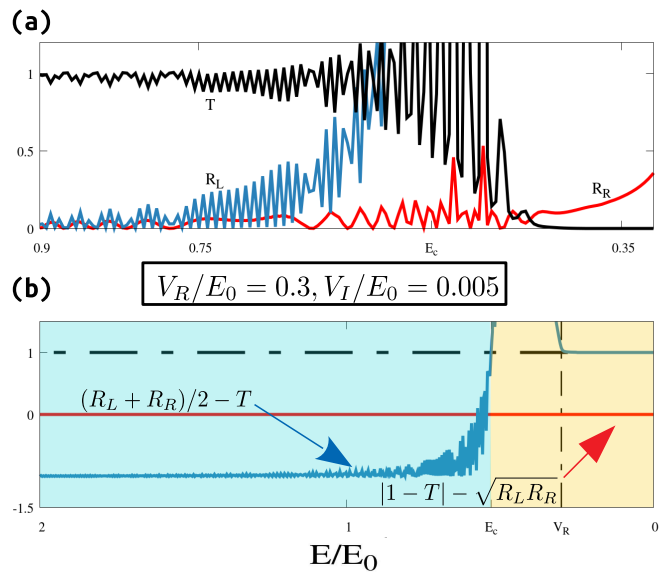}
\caption{{\bf (a)} For $L=0.5$ $\mu m$, $V_R/E_0=0.3, V_I/E_0=0.005$, $R_L$ (blue), $R_R$ (red) and $T$ (black) are plotted versus dimensionless energy $E/E_0$, where $E_0=1~eV$. The anisotropic transmission resonances (ATR), are shown, where both $R_R=0$, $T=1$ and $R_L=0$, $T=1$ are obtained for varying energies. {\bf (b)} The pseudo-unitarity condition $\vert 1 -T \vert - \sqrt{R_R R_L}=0$ is shown in red, whereas the blue curve displays a measure for $\cal PT$-symmetry; when $[R_L+R_R]/2-T<1$ $\cal PT$-symmetry holds; $[R_L+R_R]/2-T=1$ is the spontaneous symmetry breaking (SSB) point and beyond that point for $[R_L+R_R]/2-T>1$ $\cal PT$-broken phase onsets. The black dashed line displays the SSB point, that separates the $\cal PT$-symmetric (blue) and $\cal PT$-broken (yellow) phases.}
\label{check}
\end{figure}

\begin{eqnarray}
\vert 1 -T \vert = \sqrt{R_R R_L}.
\label{pseudo}
\end{eqnarray}
After applying the boundary conditions, we obtain $r_L = N_L/D_L$ and $t_R = N_T/D_T $ where,

\begin{eqnarray}
\label{long}
N_L &=& [i \, \Omega_1 sin\Lambda +\Omega_0^{\ast} cos\Lambda] [cos\Lambda^{\ast}- i \, \Omega_0^{\ast}  sin\Lambda^{\ast} ] \nonumber \\
   &+& [i \, sin\Lambda^{\ast}- \Omega_0^{\ast} cos\Lambda^{\ast} ]  [ cos\Lambda + i \, \Omega_0 sin\Lambda  ], \nonumber \\   
D_L &=& [cos\Lambda- i \, \Omega_0  sin\Lambda ] [\Omega_0^{\ast} cos\Lambda^{\ast}-i \, sin\Lambda^{\ast} ]   \\
   &+& [\Omega_0^{\ast} cos\Lambda- i \, \Omega_1  sin\Lambda ]  [ cos\Lambda^{\ast}- i \, \Omega_0^{\ast}  sin\Lambda^{\ast}  ], \nonumber \\  
N_T &=& [ cos\Lambda- i \, \Omega_0  sin\Lambda ]r_L + cos\Lambda + i \, \Omega_0  sin\Lambda, \nonumber \\
D_T &=& cos \Lambda^{\ast}- i \, \Omega_0^{\ast}  sin\Lambda^{\ast}. \nonumber
\end{eqnarray}

Here, $\Lambda\equiv k_1 L/2$ , $\Omega_0\equiv k_0/k_1$, $\Omega_1\equiv k_1/k^{\ast}_1$. From those, $r_R$ and $t_L$ can be found by exploiting the symmetry: when $k_1\mapsto k^{\ast}_1$, then $r_L\mapsto r_R$ and $t_R\mapsto t_L$. Moreover, due to reciprocity in transmission, $t_L=t_R$, so we can drop the left/right subscript, and write it as $t$. Fig.~\ref{check}(a) show $R_L, R_R$, and $T$ for varying $E$. Analogous to the $\cal PT$-symmetric optics, ATRs, where either $T=1$, $R_L=0$ or $T=1$, $R_R=0$ \cite{Ge} are also achievable in our system and can be seen in Fig.~\ref{check}(a). An important result displayed in Fig.~\ref{check}(a) is that it is possible to obtain both consecutive ATRs from left/right by fine-tuning the energy as well as separated regions of left/right ATRs, still accessible via varying the energy, which is advantageous for device applications. The validity of pseudo-unitarity condition $\vert 1 -T \vert - \sqrt{R_R R_L}=0$ is displayed in Fig.~\ref{check}(b). Another important feature shown in Fig.~\ref{check}(b) is a measure for the spontaneous symmetry breaking (SSB) point where the $\cal PT$-breaking transition onsets. The measure is given in terms of reflectances and transmittance where the SSB point is given by $[R_L+R_R]/2-T=1$; separating the $\cal PT$-symmetric ($[R_L+R_R]/2-T<1$) and $\cal PT$-broken ($[R_L+R_R]/2-T>1$) phases \cite{Ge}.

\section{$\mathbb{S}$-matrix calculations}
The transfer matrix $\mathbb{M}$ and the scattering matrix $\mathbb{S}$ for our system are defined as: $\vec{\psi}^{(R)}(z) = \mathbb{M}\vec{\psi}^{(L)}(z)$ and $\vec{\psi}_{o}(z) = \mathbb{S}\vec{\psi}_{i}(z)$ with,
\begin{eqnarray}
&\vec{\psi}^{R,L}(z)&=[\psi^{R,L}_{i,\uparrow}(z),\psi^{R,L}_{o,\uparrow}(z),\psi^{R,L}_{i,\downarrow}(z),\psi^{R,L}_{o,\downarrow}(z)]^T, \nonumber \\
\\
&\vec{\psi}_{i,o}(z)&=[\psi^{L}_{i,o,\uparrow}(z),\psi^{R}_{i,o,\uparrow}(z),\psi^{L}_{i,o,\downarrow}(z),\psi^{R}_{i,o,\downarrow}(z)]^T.  \nonumber
\end{eqnarray}
All 16 possible combinations of $L_i$, $R_i$, and $M$ where $i=0,1,2$ yields three different eigenvalue spectra. The $\mathbb{S}$-matrices for different combinations yielding the same eigenvalue spectra are connected by unitary transformations thus do not display new physics, so it is sufficient to consider these three cases. Let us first describe how the SFs affect the structure of the $\mathbb{S}$-matrices. When no SFs are inserted, the two-by-two diagonal blocks for each spin in the $\mathbb{S}$-matrix are uncoupled. Mathematically, the effect of SFs is to mix the definite spin parts of the $\mathbb{S}$-matrices. To give a full recipe for $t$: if there are even number of any of the $F_0$ and $F_1$ components and/or any number of $F_2$ components inserted, position of $t$ is unchanged, whereas if there are odd number of any of the $F_0$ and $F_1$ components and any number of $F_2$ components inserted, position of $t$ is shifted to the opposite spin entry in the $\mathbb{S}$-matrix; for $r_R/r_L$, if there is an $R_1/L_1$ component inserted, position of $r_R/r_L$ is unchanged, on the other hand if an $R_0/L_0$ or $R_2/L_2$ component is inserted, position of $r_R/r_L$ is shifted to the opposite spin entry in the $\mathbb{S}$-matrix. The effect of SFs on the $\mathbb{S}$-matrices can be fully understood by investigating Fig.~\ref{structure}(c) and Table~\ref{prop}. We will go over these effects below for the nontrivial cases, case 2 and case 3.

Going back to the three cases, the first one (case 1) is the most trivial case where the spin degrees of freedom are uncoupled due to the absence of SFs. 
\begin{ceqn}
\begin{align}
\mathbb{S}^{(1)} = \begin{pmatrix} r_R & t & 0 & 0 \cr t & r_L & 0 & 0 \cr 0 & 0 & r_R & t \cr 0 & 0 & t & r_L \end{pmatrix}.
\end{align}
\end{ceqn}
yielding the same eigenvalues twice since spin degrees of freedom are uncoupled:
\begin{ceqn}
\begin{align}
\lambda^{(1)}_{1,2} = \frac{1}{2} \Big \{ (r_R+r_L) \pm \sqrt{(r_R-r_L)^2 +4t^2} \Big \}.
\end{align}
\end{ceqn}

Second case (case 2) displays coupling of spin degrees of freedom, yet no mixed state can be achieved. As an example, case 2 is obtained when $L_0MR_0$ configuration is used, where $M$ denotes the spin-independent $\cal PT$-symmetric potential component. Using the recipe described above, even number of $F_0$ components leave the placement of $t$ unchanged in the $\mathbb{S}$-matrix and placement of $R_0/L_0$ components shifts the $r_R/r_L$ entry to the opposite spin in the $\mathbb{S}$-matrix.

\begin{ceqn}
\begin{align}
\mathbb{S}^{(2)} = \begin{pmatrix} 0 & t & r_R & 0 \cr t & 0 & 0 & r_L \cr r_R & 0 & 0 & t \cr 0 & r_L & t & 0 \end{pmatrix}.
\end{align}
\end{ceqn}

\begin{figure} [t]
\includegraphics[scale=0.95]{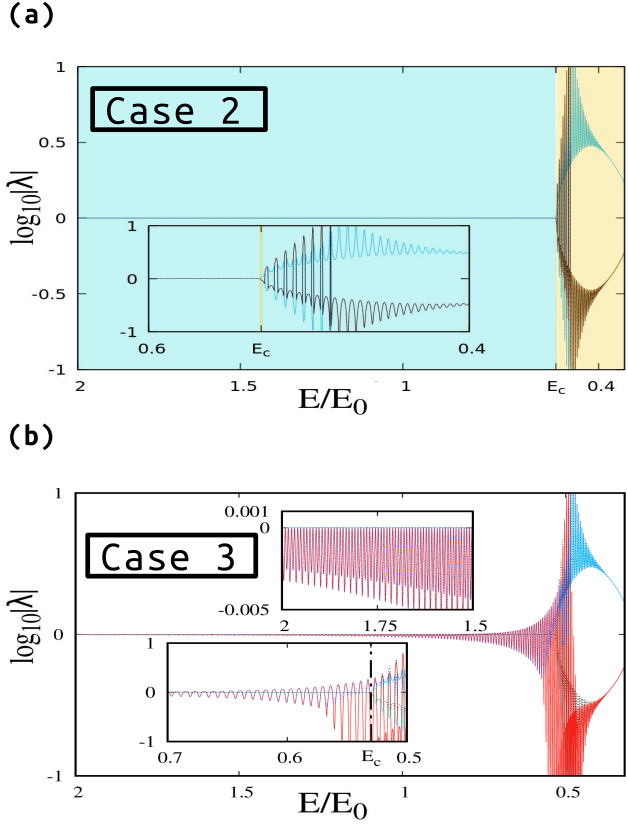}
\caption{{\bf (a)} The magnitudes of the eigenvalues of $\mathbb{S}^{(2)}$, corresponding to case 2, versus the dimensionless energy scaled with $E_0=1$ $eV$ are shown in $log_{10}$ scale for $V_R/E_0=0.3$, $V_I/E_0=0.005$. The eigenvalues of $\mathbb{S}^{(2)}$ do not display phase mixing and the $\cal PT$-symmetric (blue) and $\cal PT$-broken (yellow) eigenvalues exist in separate $E$ values. {\bf (b)} For case 3, the first set of eigenvalues $\lambda^{(3)}_{1,2}$ which are $\cal PT$-symmetric before the SSB point are displayed in teal and black, whereas the second set, $\lambda^{(3)}_{3,4}$ displayed in red and blue, which are $\cal PT$-broken even far before hitting the SSB point, as shown in two insets.}
\label{eigen}
\end{figure}

Four eigenvalues for this configuration are given as:
\begin{ceqn}
\begin{align}
\lambda^{(2)}_{1-4} = \frac{1}{2} \Big \{ \pm (r_R+r_L) \pm \sqrt{(r_R-r_L)^2 +4t^2} \Big \}.
\end{align}
\end{ceqn} 

The last case (case 3) is the most interesting case where the mixed state of eigenvalues, i.e. coexistence of  $\cal PT$-symmetric and $\cal PT$-broken eigenvalues is realized. Let us investigate the configuration $L_0M$ as an example for this case. Since we have odd number of $F_0$ components, $t$ is shifted to the opposite spin entry in the $\mathbb{S}$-matrix and also that component being $L_0$, $r_L$ is also shifted to the opposite spin entry in the $\mathbb{S}$-matrix.

\begin{figure} [t]
\includegraphics[scale=0.75]{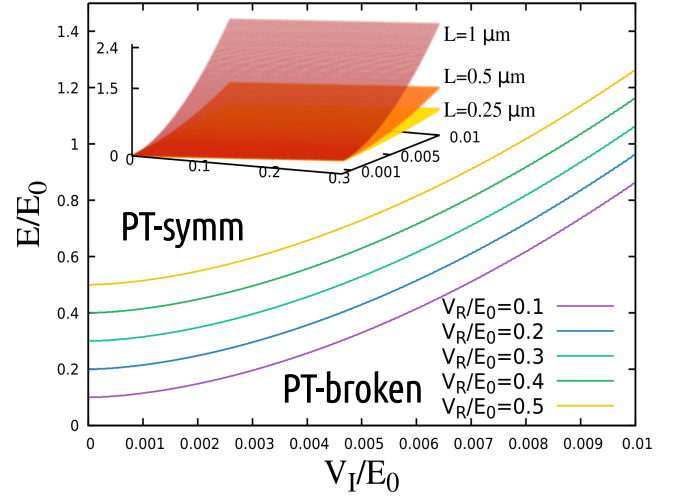}
\caption{The curves for the case $L=0.5$ microns separating the $\cal PT$-symmetric upper and $\cal PT$-broken lower parts for varying $V_R/E_0$ values and the manifold (inset) of SSB, which separates the $\cal PT$-symmetric upper and $\cal PT$-broken lower parts for the given set of variables for $L=0.25$ (yellow), $0.5$ (orange) and $1$ microns (red) are shown.}
\label{ssb}
\end{figure}

\begin{ceqn}
\begin{align}
\mathbb{S}^{(3)} = \begin{pmatrix} r_R & 0 & 0 & t \cr 0 & 0 & t & r_L \cr 0 & t & r_R & 0 \cr t & r_L & 0 & 0 \end{pmatrix}.
\label{eq:Eq7}
\end{align}
\end{ceqn}
This configuration yields the following eigenvalues: 
\begin{subequations}
\begin{ceqn}
\begin{align}
\lambda^{(3)}_{1,2} = \frac{1}{2} \Big \{  (r_R+r_L) \pm \sqrt{(r_R-r_L)^2 +4t^2} \Big \}, \\
\lambda^{(3)}_{3,4} = \frac{1}{2} \Big \{  (r_R-r_L) \pm \sqrt{(r_R+r_L)^2 +4t^2} \Big \}.
\end{align}
\end{ceqn}
\end{subequations}

The eigenvalues of the $\mathbb{S}$-matrices are unimodular for the $\cal PT$-symmetric phase. When $\cal PT$-symmetry is broken, they become reciprocal pairs.

\section{Results}

\begin{table} 
\caption{Different configurations depicted in Fig.~\ref{structure}(c).}
\begin{tabular}{ l| a | d | a | d }
 \hline \hline
 \rowcolor{camel}
 configuration & eigenvalues & components &phase-mix & case \\ \hline
 ${\bf M}$ & 2 & 1 & \xmark & 1 \\ \hline 
 {$\bf L_0$}{$\bf M$}{$\bf R_0$} & 4 & 3 & \xmark & 2  \\ \hline 
 {$\bf L_0$}{$\bf M$} or {$\bf M$}{$\bf R_0$} & 4 & 2 & \cmark & 3 \\ \hline
 {$\bf L_1$}{$\bf M$}{$\bf R_1$} & 4 & 3 & \xmark & 2 \\ \hline
 {$\bf L_2$}{$\bf M$}{$\bf R_2$} & 4 & 3 & \xmark & 2 \\ \hline
 {$\bf L_1$}{$\bf M$} or {$\bf M$}{$\bf R_1$} & 4 & 2 & \xmark & 2 \\ \hline
 {$\bf L_2$}{$\bf M$} or {$\bf M$}{$\bf R_2$} & 4 & 2 & \cmark & 3 \\  \hline
 {$\bf L_0$}{$\bf M$}{$\bf R_1$} or {$\bf L_1$}{$\bf M$}{$\bf R_0$} & 4 & 3 & \cmark & 3 \\ \hline
 {$\bf L_0$}{$\bf M$}{$\bf R_2$} or {$\bf L_2$}{$\bf M$}{$\bf R_0$} & 4 & 3 & \xmark & 2 \\ \hline  
 {$\bf L_1$}{$\bf M$}{$\bf R_2$} or {$\bf L_2$}{$\bf M$}{$\bf R_1$} & 4 & 3 & \cmark & 3 \\ \hline \hline  
\end{tabular}
\label{prop}
\end{table} 

Fig.~\ref{eigen} displays the $\cal PT$-symmetric and $\cal PT$-broken eigenvalues for a structure with realistic parameters of $L=0.5$ microns, $V_R/E_0=0.3$ and $V_I/E_0=0.005$ with varying energy spectra (scaled with $E_0=1$ $eV$) for case 2 and case 3. As mentioned hereinabove, for the eigenvalues of $\mathbb{S}^{(2)}$, i.e. case 2, no phase mixing can be obtained for any $E$, as shown in Fig.~\ref{eigen}(a). For case 3, the two different sets of eigenvalues corresponding to $\mathbb{S}^{(3)}$ have a different behavior as shown in Fig.~\ref{eigen}(b): The first set of eigenvalues $\lambda^{(3)}_{1,2}$ are $\cal PT$-symmetric before reaching the spontaneous symmetry breaking (SSB) point or in other words the critical energy $E_c\simeq0.53$. Those eigenvalues are displayed in teal and black. The other set $\lambda^{(3)}_{3,4}$ are $\cal PT$-broken even far before reaching the SSB point. It is important to note that the magnitudes of $\lambda^{(3)}_{1,2}$ and $\lambda^{(2)}_{1-4}$ are equal, pointing that $\lambda^{(3)}_{3,4}$ are causing the mixing of $\cal PT$-symmetric and $\cal PT$-broken eigenvalues. The second set $\lambda^{(3)}_{3,4}$ are displayed in red and blue and the magnitudes of all four eigenvalues are plotted in $log_{10}$ scale. Two insets of Fig.~\ref{eigen}(b) display that even for large values of $E/E_0$ before reaching $E_c$, the second set $\lambda^{(3)}_{3,4}$ are $\cal PT$-broken, hence a broadband energy spectrum for phase mixing can be achieved.

Another significant result of this letter is displayed in Fig.~\ref{ssb}, in which the curves for the case $L=0.5$ microns separating the $\cal PT$-symmetric upper and $\cal PT$-broken lower parts for six different $V_R/E_0$ values are shown. In the inset, the SSB manifold (with $V_R$ dependence now explicitly shown together with $V_I$) is plotted for $L=0.25,0.5,1$ microns with yellow orange and red, respectively, where $L$ is the size of the component $M$. In the figure (and inset), the upper part of the curves (manifolds) corresponds to the set of variables that display $\cal PT$-symmetric eigenvalues for the $\mathbb{S}$-matrix and vice versa for the lower part.

Table. \ref{prop} summarizes all possible 16 configurations. The minimum number of components to achieve mixing is two. It is possible to achieve phase mixing also with three components. More interestingly, the configurations $L_1 M$ and $M R_1$ give no phase mixing, signifying the importance of the reflectance properties of the SFs over the transmittance properties for obtaining phase mixing. 

Finally, as we demonstrated in the previous section, ATRs with extended features are accessible in our system. Fig.~\ref{check}(a) displays that, consecutive ATRs from left/right by fine-tuning the energy as well as separated regions of left/right ATRs are accessible.

\section{Discussion and Conclusions}

The theoretical scheme that we proposed is promising for a variety of applications: The first one is in the field of spintronics. Similar to the applications of $\cal PT$-symmetry in optics, uni-directional invisibility, lasing, and other enthusing phenomena can be achieved in quantum mechanical systems. By introducing spin-$\frac{1}{2}$ particles and suggesting a structure that allows phase mixing demonstrated in this letter, it would be possible to achieve all of these properties selectively in spintronics based devices. Moreover, in our theoretical studies, we obtained switching between left/right ATRs both in a close neighborhood of energy values as well as in a relatively separated energy region, for the same parameter space, which is promising for non-reciprocal applications. The incoming energy of the fermionic species depend on their chemical potential which can be controlled through an applied gate potential, hence creating a controllable diode-like device. Secondly, in semiconductor-based devices, such as coupled quantum wells, obtaining mixed phase as suggested in this work would allow multi-functionality within a broadband spectrum. Lastly, the geometry in Fig.~\ref{structure} can be extended to study a spin-dependent $\cal PT$-symmetric potential. This can be relevant in $\cal PT$-symmetric topological surface states with a strong spin-orbit coupling that has gained attention recently \cite{PT_symm_TSS}. It is important to state that all of the parameters that we used in this letter have realistic values for the above-mentioned applications.

To conclude, we have theoretically demonstrated the mixing of $\cal PT$-symmetric and $\cal PT$-broken eigenvalues of the $\mathbb{S}$-matrix describing the one-dimensional spin-$\frac{1}{2}$ scattering problem from a spin-independent complex $\cal PT$-symmetric potential, for broadband energy spectra. We studied 16 different configurations obtained by combining SFs and $M$ in all possible ways and categorized these in terms of their phase-mixing properties. Moreover, we discussed the analogies with $\cal PT$-symmetric optics and theoretically demonstrated the existence of ATRs with extended features in our scheme. We believe that our results will be promising for spintronics applications, semiconductor-based devices, and can contribute to the further understanding of topological surface states.

\acknowledgments
One of the authors (E.O.) acknowledges partial support from the Turkish Academy of Sciences. This manuscript is originally published in EPL as {\it Europhys. Lett. 131 11001 (2020).}

\end{document}